%% file: arXiv_main.tex
\documentclass{article}

\usepackage{amsmath}
\usepackage{amssymb}
\usepackage{graphicx}
\usepackage{dcolumn}
\usepackage{bm}
\usepackage{hyperref}
\usepackage{color}
\usepackage[a4paper, total={6in, 8in}]{geometry}

\title{Emergence of gravity from quantum field theory in triangulated spacetime and the QFT vector model}
\author{Matti Raasakka\\ \small{\tt{matti.raasakka@aalto.fi}} \\\small{Micro and Quantum Systems group, Aalto University, Espoo, Finland}}

\begin{document}

\maketitle

\begin{abstract}
We formulate quantum field theory in triangulated spacetimes using compositional quantum field theory and tensor network methods. We show that, in the case of free massive scalar field theory in two-dimensional Lorentzian spacetime, an effective gravitational action emerges from the quantum field theory amplitudes, similarly to Sakharov's induced gravity approach. The generalization to higher dimensions and other models is also discussed, but the details are left for future work. Our results lead us to propose a new approach to the unification of quantum field theory with gravity, the \emph{QFT vector model}, which combines insights and techniques from various current approaches to quantum gravity such as causal dynamical triangulations, random tensor models, group field theory, emergent gravity and holography. The key idea behind the approach is that, if indeed matter QFT induces effective quantum gravitational amplitudes, then we may not need to include gravitational dynamics explicitly into the fundamental model in order to obtain a unified theory of quantum matter and gravity. 
\end{abstract}

\section{Introduction}
Quantum gravity has puzzled physicists for nearly a century by now. Despite a lot of progress on many different approaches \cite{Oriti_book}, the final solution to merging quantum field theory (QFT) with general relativity (GR) remains elusive. The perturbative nonrenormalizability of the QFT model that arises as a direct quantization of GR has prevented the formulation of a fundamental theory of quantum gravity in the same way as for the other fundamental forces of nature \cite{Burgess04}. However, this nonrenormalizability may also be taken as a hint towards the effective nature of gravitational phenomena. Indeed, the fundamental nature of gravitational interactions and even spacetime itself has been called into question by various results along the way \cite{Jacobson95,Verlinde11}. As early as 1967 Andrei Sakharov showed that gravitational interactions can emerge as an effective phenomenon from the vacuum fluctuations of matter QFT on curved spacetimes \cite{Sakharov68} --- so-called \emph{induced gravity}. This turns out to be a very general result, since the effective action that arises in this way consists necessarily of geometric invariants, which in the lowest orders match up exactly in form with the Einstein--Hilbert action for GR \cite{Visser02}. We will follow Sakharov's lead in this paper as we uncover a similar phenomenon in QFT in triangulated spacetimes instead of continuous ones. The concrete results and calculations concern the case of free massive scalar field theory in a triangulated 2d Lorentzian spacetime, but we also discuss how they can be generalized to higher dimensions and other QFT models. Our results lead us to propose a new approach to the unification of QFT with gravity --- the \emph{QFT vector model} --- which combines insights and techniques from various current approaches to quantum gravity such as causal dynamical triangulations, random tensor models, group field theory, emergent gravity and holography. The key idea behind the approach is that, if indeed matter QFT induces effective quantum gravitational amplitudes, then we may not need to include gravitational dynamics explicitly into the fundamental model in order to obtain a unified theory of quantum matter and gravity. In this paper we fill in some of the details and expand on the discussion in the conference paper \cite{Raasakka26}. In particular, the case of 2d free massive scalar QFT is analyzed in detail, and the emergence of an effective gravitational action shown explicitly.

Triangulated spacetimes are considered in many of the current approaches to quantum gravity, such as causal dynamical triangulations (CDT) \cite{Ambjorn01}, spin foams \cite{Perez13}, group field theory \cite{Oriti09} and random tensor models \cite{Francesco95,Gurau2024}. In addition to the potential ill-definedness of spacetime geometry below the Planck scale $l_P = \sqrt{\hbar G/c^3} \sim 10^{-35}$ meters, a practical reason for introducing the triangulation is to make models well-behaved by imposing a high-energy cutoff to the geometric degrees of freedom. Moreover, since the work of Regge \cite{Regge61}, it has been understood that GR can be formulated in a neat coordinate-free way by using triangulated manifolds, thus sidestepping completely the difficult problem of imposing general covariance under coordinate transformations. Curvature --- concentrated on codimension-$2$ faces of the triangulation --- is quantified by deficit angles $\theta_f$, i.e., the amount of rotation (or, more generally, Lorentz transformation) a gyroscope experiences when going once around the face $f$. The Einstein--Hilbert action for GR then takes an intuitive geometric form $S_\text{EH}[\Delta] = (8\pi G)^{-1}\sum_f V_f \theta_f$ for a triangulated manifold $\Delta$, where the sum runs over all the faces $f\in\Delta$, $G$ is the gravitational constant and $V_f$ the volume of the face $f$ \cite{Sorkin75}. (The cosmological constant term, which we neglect in this main article, is discussed in Appendix \ref{app:regge}.) This is taken advantage of, e.g., in CDT and random tensor models \cite{Francesco95,Gurau2024} to formulate a quantum gravitational path integral over spacetime geometries, where each triangulation is weighted by the amplitude $e^{iS_\text{EH}[\Delta]/\hbar}$. For us the triangulation will also serve another purpose: Since the gravitational constant $G$ emerges in our approach as an effective parameter, whose numerical value is computed from the more fundamental description, the discretization scale $l_\Delta$ of the triangulation (instead of the Planck length $l_P$) determines an absolute length scale.

In the approach we take in this paper the field theory degrees of freedom are not discretized on a lattice, but the continuous fields live in a triangulated piecewise-flat continuum spacetime. Such an approach is taken in order to avoid challenges in recovering the known continuum physics from the discretized theory.\footnote{It is however an interesting open question whether similar results to the ones in this paper can be achieved also in a lattice approach to QFT, such as recently studied e.g.\ in \cite{Evnin26}. We do not see why it would not be possible if the model can be formulated in a similar manner to the continuum case, but due to the above motivations we have not pursued such an approach in this work.} In formulating the theoretical framework, we draw inspiration from the general boundary formulation of QFT (GBQFT) introduced by Oeckl \cite{Oeckl03b,Oeckl06,Oeckl08} and its later expansion, the compositional QFT (CQFT) approach \cite{Oeckl24}. GBQFT generalizes standard formulations of QFT by associating state spaces and probability amplitudes to arbitrary (possibly bounded and finite) spacetime regions. Furthermore, CQFT provides an axiomatic framework for gluing spacetime regions and their associated probability amplitudes together to obtain amplitudes for composite regions. (See \cite{Oeckl06,Oeckl08,Oeckl24} for further details.) We will use CQFT \cite{Oeckl24} and tensor network methods \cite{Orus19} to formulate our model first on a single flat $d$-simplex --- the $d$-dimensional generalization of a triangle --- and then glue simplices together to obtain QFT amplitudes for more complicated triangulated manifolds. This will give us a general framework for constructing QFT models on piecewise-flat triangulated spacetimes. Importantly, the framework also allows to describe physics inside a spatiotemporally bounded spacetime region. Having the possibility to impose classical geometric boundary data for the bounded region helps with the operational interpretation of the model --- especially when we move on to quantum gravity. The boundary can be used to keep track of classical observer time, thus sidestepping the problem of time(lessness) in quantum gravity \cite{Oeckl03}. This can also be thought of as degrees of freedom on the boundary providing a (quantum) reference frame for the bulk system \cite{Kabel23}.

Tensor networks were originally developed as an effective description for ground states of many-body quantum systems, but have since found applications in many different contexts \cite{Orus19}. Our approach draws inspiration from previous works, where tensor networks have been related to spacetime path integrals \cite{May17,Milsted18,Yang19,Pedraza22} and holography \cite{Swingle12B,Pastawski15}. The basic insight here is that, when quantum amplitudes are expressed as a tensor network contraction, the network structure is reflected in the correlation structure of the amplitudes. On the other hand, for low-energy QFT states the correlation structure closely reflects the geometric structure of the background spacetime due to locality --- e.g., subsystems with small spatial separation being more correlated than subsystems far apart --- thus linking the tensor network structure to spacetime geometry. Following ideas from holography and other works relating entanglement to spacetime structure \cite{VanRaamsdonk10,Bianchi12,Maldacena13,Cao17,Swingle18}, we postulate that fundamentally spacetime structure \emph{is} just an effective description of the correlation structure of the low-energy states of the system \cite{Raasakka17}. This then implies that \emph{outside} the low-energy regime the QFT boundary amplitudes may not reflect a unique bulk geometry but may be expressed as a linear combination of many different tensor network contraction patterns, leading to a kind of superposition of bulk geometries.\footnote{Notice that since the framework does not have fundamental geometric observables, but the bulk geometry is just an effective description of the boundary correlation structure, the different bulk geometries are not in any fundamental sense orthogonal in the Hilbert space, even though some approximate notion of orthogonality may be recoverable in the macroscopic limit.} We will employ random tensor model techniques to write down a partition function, which generates the sum over all possible QFT tensor network amplitudes compatible with a given boundary data, giving rise to what we call the QFT vector model.

\section{Tensor networks for QFT amplitudes in triangulated spacetimes}
Let us begin the concrete construction of the framework by considering a single $d$-simplex $T$ of linear size $l_\Delta$ embedded in a flat $d$-dimensional Lorentzian spacetime. Following the principles of compositional QFT \cite{Oeckl24}, we associate a quantum state space $\mathcal{H}_i$ to each boundary $(d-1)$-simplex, or \emph{face}, $F_i\subset \partial T$, where $i=0,\ldots,d$.\footnote{We note that the boundary state spaces may be Krein spaces (instead of Hilbert spaces) for a model containing fermionic degrees of freedom \cite{Oeckl24}.} Let us assume for simplicity that our model consists of a single scalar field, although the construction can be generalized to arbitrary field content. In this case, $\mathcal{H}_i$ can be constructed as a Fock space by first fixing a basis of functions $\zeta_k$ in $L^2(F_i,\mathbb{C})$ ($k\in\mathbb{N}$), the \emph{modes} of the field on the face $F_i$. Then $\mathcal{H}_i$ is linearly spanned by the number states
\begin{align}
    |{\bf n}_i\rangle := |n^{(i)}_1,\ldots,n^{(i)}_K\rangle \in \mathcal{H}_i \,,
\end{align}
where $n_k$ is the number of excitations in the mode corresponding to the basis function $\zeta_k$ and $K\in\mathbb{N}$. The \emph{kinematical} state space of the total boundary $\partial T$ is obtained as the symmetrized tensor product $\mathcal{H}_{\partial T}^\text{kin} = S(\otimes_i \mathcal{H}_i)$, where $S$ represents the symmetrization of the boundary excitations according to the statistics of the QFT model. The kinematical boundary state space contains also states, which are not compatible with the dynamics of the model, but these will get removed in the next step.

Next we introduce the amplitude map, which encodes the dynamics of the model by associating probability amplitudes to boundary states \cite{Oeckl24}. Due to its linearity the amplitude map is an element $\langle A|$ in the dual space $(\mathcal{H}_{\partial T}^\text{kin})^*$, and the amplitude for the boundary state $| {\bf n}_0 \ldots {\bf n}_d\rangle \equiv S(\otimes_i |{\bf n}_i\rangle) \in \mathcal{H}_{\partial T}^\text{kin}$ is given by $\langle A | {\bf n}_0 \ldots {\bf n}_d\rangle$. Concretely, the amplitude can be computed by taking a copy of the corresponding mode function for each boundary excitation, and then integrating over their product using the $n$-point function $G_n$ of the QFT model as the integral kernel. In mathematical notation,
\begin{align}
	\langle A | (\otimes_{i=0}^d | n_1^{(i)},\ldots, n_{K_i}^{(i)} \rangle )
    = \left[ \prod_{i=0}^d \prod_{k_i=1}^{K_i} \prod_{m_{k_i}^{(i)}=1}^{n_{k_i}^{(i)}} \int_{F_i} \text{d}^{d-1}x_{k_i,m_{k_i}^{(i)}}^{(i)} \zeta_{k_i}(x_{k_i,m_{k_i}^{(i)}}^{(i)}) \right] G_n(x_{1,1}^{(1)}, \ldots,x_{K_I,m_{K_I}}^{(I)}) \,,
\end{align}
where $n=\sum_i\sum_{k_i} n_{k_i}^{(i)}$ is the total number of excitations on the faces. For a flat $d$-simplex, $G_n$ will be the flat spacetime $n$-point function, since inside the simplex the physics should be that of flat spacetime according to the equivalence principle.\footnote{While our choice here for the boundary amplitudes is physically motivated by the equivalence principle, other quantization schemes specifically developed for bounded spacetime regions (such as e.g.\ in \cite{Oeckl13}) can also be considered, possibly leading to different amplitudes.} This choice will also guarantee the preservation of local Lorentz symmetry. The \emph{physical} boundary state space is then obtained as $\mathcal{H}_{\partial T}^\text{phys} = P_{\text{ker}A}^{\perp} \mathcal{H}_{\partial T}^\text{kin}$,
where $P_{\text{ker}A}^{\perp}$ is the projection onto the orthogonal subspace to the kernel of the amplitude map. This projection gets rid of any unphysical boundary states, which do not respect the dynamics of the model.

\begin{figure}
    \centering
    \def\svgwidth{0.8\linewidth}
	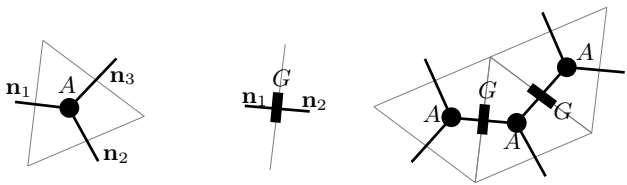
    \caption{Left: The graphical representation of the amplitude tensor $A_{\mathbf{n}_1\mathbf{n}_2\mathbf{n}_3}$ corresponding to a triangle (gray) as a circle with three legs. Center: The graphical representation of a gluing tensor $G_{\mathbf{n}_1\mathbf{n}_2}$ corresponding to an internal face (gray) of a triangulation as a rectangle with two legs. Right: The graphical representation of a contraction of three amplitude tensors, mediated by gluing tensors for the internal faces. Boundary legs carry free indices, which are not marked in the figure.}
    \label{fig:tensor}
\end{figure}

We can now think of the amplitudes $\langle A | {\bf n}_1 \ldots {\bf n}_{d+1}\rangle$ in the Fock basis as tensor components of an infinite-dimensional `spacetime tensor', 
\begin{align}
    A_{{\bf n}_1 \ldots {\bf n}_{d+1}} \equiv \langle A | {\bf n}_1 \ldots {\bf n}_{d+1}\rangle \,.
\end{align}
which we will call the \emph{amplitude tensor} in our model. We can then obtain amplitudes for any triangulated manifold by contracting a number of amplitude tensors according to the geometric structure of the triangulation, because the tensor contraction corresponds to a sum over the Fock basis on a particular face of the triangulation. However, when gluing the amplitude tensors together, we have to account for the orientation of the boundaries by introducing the gluing tensor $G_{{\bf n}_1 {\bf n}_{2}}$, which maps the boundary modes ${\bf n}_1$ on one boundary simplex to the modes ${\bf n}_2$ on the other in a way that implements the change in orientation. (See Fig.\ \ref{fig:tensor} for a 2-dimensional example.)

In many cases it may be necessary (or otherwise preferable) to consider several different kinds of simplices, or even other polytopes. The generalization of the above framework is straightforward: We have several different amplitude tensors corresponding to the different geometric shapes, which can be contracted along matching faces.

\section{Effective gravitational action from free scalar QFT in 2d triangulated spacetime}
We now specialize to the free massive scalar field model in 2d Lorentzian spacetime, allowing us to carry out concrete calculations quite easily. We will discuss the generalization to other models and higher dimensions after presenting the results in this simple case.

\begin{figure}
    \centering
    \def\svgwidth{0.8\linewidth}
	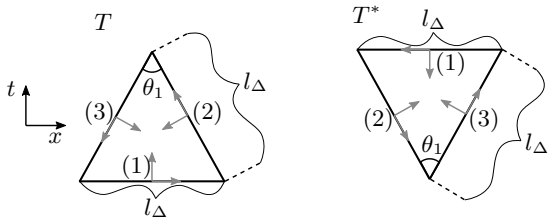
    \caption{Two oppositely oriented Lorentzian equilateral triangles $T$ and $T^*$ with one spacelike edge (1) and two timelike edges (2) and (3) of proper length/time $l_\Delta$. Positive directions on the edges are indicated by grey arrows.}
    \label{fig:lorentziantriangles}
\end{figure}

\begin{figure}
    \centering
    \def\svgwidth{0.8\linewidth}
	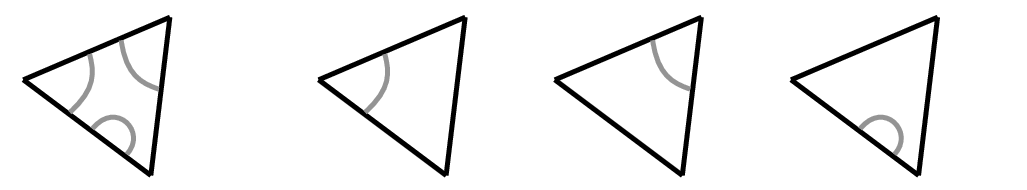
    \caption{The decomposition of a multi-particle process amplitude into a product of single-particle process amplitudes.}
    \label{fig:decomp}
\end{figure}

In two dimensions a triangulated manifold consists of 2-simplices, i.e., triangles. Similarly to 2d CDT \cite{Loll15}, we will consider two oppositely oriented equilateral triangles of linear size $l_\Delta$, as depicted in Fig.\ \ref{fig:lorentziantriangles}, from which more complicated triangulated manifolds can be constructed. Functions of the form $\zeta_k(s) = e^{2\pi i ks / l_\Delta}$, $k\in\mathbb{Z}$, form a basis in $L^2([-\frac{l_\Delta}{2},\frac{l_\Delta}{2}],\mathbb{C})$ as they satisfy
\begin{align}
    (\zeta_k,\zeta_l) = \int \text{d}s\ \zeta_k(s) \overline{\zeta_l(s)} = l_\Delta^{-1}\delta_{kl} \,.
\end{align}
State spaces on each edge of the triangle are constructed as Fock spaces using $\zeta_k$ as a basis of field modes, as described earlier. The gluing tensor $G$ maps the modes as $k \mapsto -k$, when the edges to be identified are oppositely oriented.

In the free model the $n$-point function decomposes into a sum of products of $2$-point functions due to Wick's theorem \cite{Huang10}. Thus, an amplitude with $2n$ boundary excitations can be decomposed into a sum of products of $n$ amplitudes with two boundary excitations each. The amplitude for any odd number of boundary excitations is zero. (See Fig.\ \ref{fig:decomp} for a graphical illustration of the decomposition of one such term.) The amplitude for an excitation in mode $k$ on edge $i$ and an excitation in mode $l$ on edge $j$ is given by
\begin{align}
    B_{kl}^{(ij)} &= \int \text{d}s \int \text{d}s'\ \zeta_k(s) \zeta_l(s') G_2({\bf x}^{(i)}(s),{\bf x}^{(j)}(s')) \,,
\end{align}
where $G_2$ is the causal 2-point function of the model and ${\bf x}^{(i)}(s)$ is the coordinate vector for points on edge $i$ parametrized by $s\in [-\frac{l_\Delta}{2},\frac{l_\Delta}{2}]$. The amplitudes for the triangles $T$ and $T^*$ are equal, since the two triangles are mapped to each other by a spacetime reflection (i.e., a $PT$-transformation), and the 2-point function is invariant under this transformation. Let us call $B^{(ij)}$ the \emph{single-particle process matrices}. The elements $B^{(ij)}_{kl}$ can be explicitly numerically evaluated up to some mode cutoff $|k|,|l|\leq n_\text{max}$. They fall off as $O(|k|^{-1},|l|^{-1})$, allowing us to get good numerical estimates for the virtual particle contributions to the amplitudes we are interested in the following. (For numerical analysis we used the values $l_\Delta = 1$, $m = 1$ and $n_\text{max}=50$. See Appendix \ref{app:Bs} for details.) We used the SciPy Python library \cite{scipy} for numerical computations. Our Python scripts can be found at \cite{git}.

\begin{figure}
    \centering
    \def\svgwidth{0.7\linewidth}
	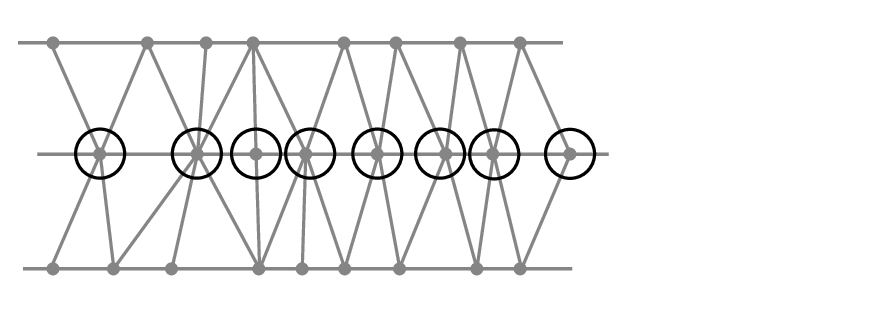
    \caption{A portion of a triangulated manifold with three unique equal-time slices at times $t$ and $t\pm\Delta t$. The leading order contribution to the effective low-energy action is given by virtual particle loops around single vertices (in black).}
    \label{fig:renorm}
\end{figure}

\begin{figure}
    \centering
    \def\svgwidth{0.4\linewidth}
	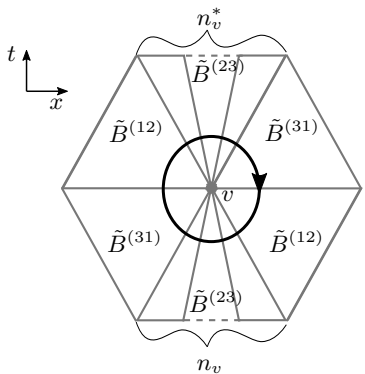
    \caption{One virtual particle loop around a single vertex $v$ of the triangulation with $n_v$ triangles $T$ to the past of $v$ and $n_v^*$ triangles $T^*$ to the future. The amplitude is obtained by contracting the single-particle process matrices with the gluing tensor according to the connectivity of the triangles.}
    \label{fig:oneloop}
\end{figure}

\begin{figure}
    \centering
    \def\svgwidth{0.6\linewidth}
	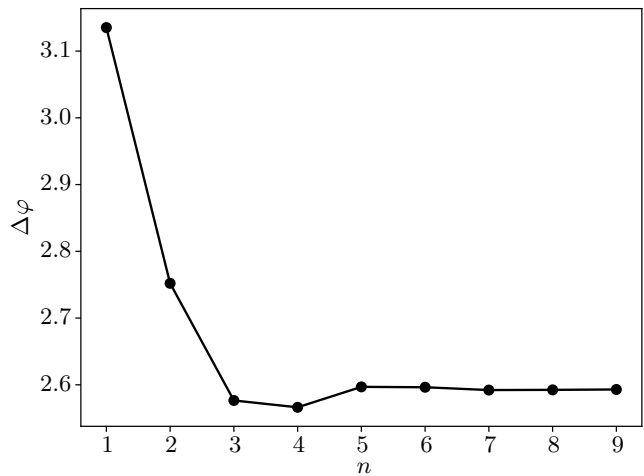
    \caption{The difference in the complex phase angle $\Delta\varphi$ of the vertex loop amplitude $A_v$ between consecutive values of $n_v$ as a function of $n_v$ with fixed $n_v^*=0$. The dependence is similar for other values of $n_v^*$. The difference $\Delta\varphi$ converges to a constant, indicating an approximately linear dependence of the phase angle on $n_v$ for $n_v>2$. Since the amplitude is symmetric with respect to $n_v$ and $n_v^*$, the same conclusion applies to $n_v^*$.}
    \label{fig:ampang}
\end{figure}

Now consider for example a portion of a triangulated 2d manifold such as illustrated in Fig.\ \ref{fig:renorm}. Such triangulations with unique equal-time slices are considered, for example, in CDT. We are interested in the amplitude between the equal-time slices at $t\pm\Delta t$, which we obtain by contracting the amplitude tensors using the gluing tensor according to the structure of the triangulation. If we are interested in the low-energy behavior of the model, we can derive the \emph{effective low-energy action} for the model by summing over high-energy degrees of freedom up to some cutoff scale $\Lambda$ \cite{Burgess04}. Since the lowest non-zero mode $\zeta_1$ on an edge has wavelength $\sim l_\Delta$, we can implement the sum over high-energy degrees of freedom by setting the triangulation length scale to $l_\Delta = 1/\Lambda$ and summing over all modes except for $k=0$. The leading order contribution to the amplitude comes from virtual particle loops around single vertices of the triangulation at the intermediate time $t$, such as illustrated in Fig.\ \ref{fig:renorm}. Let us denote by $\tilde G \equiv \delta_{k,-k}$ and $\tilde B$ the gluing and single-particle process matrices with the zero mode removed. Then, the multiplicative contribution $A_v$ to the full amplitude given by one such vertex loop around the vertex $v$, as in Fig.\ \ref{fig:oneloop}, is given by
\begin{align*}
    \text{tr}((\tilde G \tilde B^{(23)})^{n_v} \tilde G \tilde B^{(31)} \tilde G \tilde B^{(12)} (\tilde G \tilde B^{(23)})^{n_v^*} \tilde G \tilde B^{(31)} \tilde G \tilde B^{(12)}) \,,
\end{align*}
where $n_v$ is the number of $T$ triangles to the past of $v$ and $n_v^*$ is the number of $T^*$ triangles to the future of $v$.
Due to the fact that the $\tilde B^{(23)}$ matrix elements are dominated by the lowest order modes, $\tilde G \tilde B^{(23)}$ has one dominating eigenvalue $\sigma$, which is more than twice as large as the next largest. Thus, we can approximate
\begin{align}\label{eq:lin}
    A_v &\approx C \sigma^{n_v + n_v^*}\,,
\end{align}
where $C$ is some constant independent of $n_v$ and $n_v^*$. By a direct numerical evaluation we find that the complex phase of $A_v$ is closely proportional to $e^{i\Delta\varphi (n_v + n_v^*)}$, where $\Delta\varphi \approx 2.6$, with some deviation for $n_v, n_v^* < 3$. (See Fig.\ \ref{fig:ampang}.) This behavior of the phase angle can be interpreted as the gravitational Aharonov--Bohm effect \cite{Aharonov59}, where the amplitude of a virtual particle picks up an additional phase when going around a region with non-zero curvature. In our case, the approximately linear behavior of the phase angle produces an effective gravitational amplitude of the correct form $e^{iS_\text{EH}[\Delta]}$, since the number of triangles $n_v + n_v^*$ is related linearly to the deficit angle:
\begin{align}
    \theta_v = (2-n_v-n_v^*)\theta_1 \,,    
\end{align}
where $\theta_1 = \text{arccosh}(\frac{3}{2}) \approx 0.96$ is the hyperbolic angle opposite to edge $(1)$ in triangle $T$. In 2d we have $S_\text{EH}[\Delta] = - \frac{1}{8\pi G} \sum_{v\in\Delta} \theta_v$, where the summation runs over all the vertices in $\Delta$. (Notice that timelike deficit angles around spacelike faces contribute negatively to the action \cite{Sorkin75}. The volume of a vertex is set to $1$ following \cite{Ambjorn01}.) Accordingly,
\begin{align}
    e^{iS_\text{EH}[\Delta]} = \prod_v e^{\frac{i}{8\pi G}(n_v + n_v^*)\theta_1} \,,
\end{align}
which matches in form the contribution $\prod_v e^{i\Delta\varphi (n_v + n_v^*)}$ to the amplitude from the vertex loops. Equating the numerical factors gives value $G_\text{eff} = \frac{\theta_1}{8\pi\Delta\varphi} \approx 0.015$ to the effective gravitational constant.\footnote{Obviously, the exact numerical values for the effective gravitational and cosmological constants in 2d do not have any immediate physical significance, but the method for calculating them generalizes to the physically more relevant higher dimensional cases.} We have thus demonstrated the emergence of effective low-energy gravitational action from the QFT vacuum fluctuations in free scalar field model in 2d triangulated spacetime.

The deviations from strictly linear dependence of the action on the curvature in this simple model may be interpreted as higher order corrections to the Regge action in the form of a curvature-dependent gravitational constant. However, we do not yet have a precise understanding of this relation. (Higher order curvature corrections may also arise, e.g., from virtual particle loops encircling a single vertex multiple times.) It is also an open question whether other more realistic models would produce a more linear behavior. As we have seen, the degree of linearity depends essentially on the relative magnitudes of the eigenvalues of the single-particle process matrices (see Eq. \ref{eq:lin} and the preceding discussion). Other quantization schemes for bounded spacetime regions (e.g., such as considered in \cite{Oeckl13}) may also lead to different amplitudes and therefore different behavior of the effective gravitational constant. Another interesting point to consider is that when we move on to study quantum gravity, a single smooth macroscopic spacetime manifold will correspond to a very large number of different microscopic triangulations which approximate the single macroscopic geometry. E.g., for our simple 2d model we observe large deviations from linearity only for vertices missing one or two triangles, as compared to a flat 2d triangulation. Such a ‘defect’ corresponds to a huge Planck-scale amount of positive curvature around the vertex. On the other hand, if we have a small amount of positive curvature (as compared to the Planck scale) in our macroscopic spacetime, such a macroscopic configuration will be approximated by an average over many microscopic triangulations, which contain only a very few Planck-scale curvature defects. Thus, the averaged-out effective gravitational constant will also be very close to the flat spacetime value. Accordingly, the deviations from linearity would become experimentally visible only in the very extreme Planckian regimes.

Let us also emphasize that here we have considered the model purely on the level of quantum amplitudes. We showed that the effective gravitational action, along with the original matter action, appears in the exponent of the amplitudes in the (approximately) correct way. Solutions to the classical equations of motion can then be recovered as stationary points of the full action. Hence, the correct form of the quantum amplitudes guarantees also the correct classical dynamics (at least in the absence of anomalies and other such complications). That being said, the simple classical limit $\hbar\rightarrow 0$ is not sufficient to recover classical gravity, since in this limit also the quantum corrections which give rise to the effective gravitational action disappear. In particular, $G_\text{eff}\rightarrow 0$ in this limit. Instead, to recover the classical gravitational dynamics, we would need to take simultaneously the continuum limit $l_\Delta \rightarrow 0$ in such a way that the effective gravitational constant remains fixed. (See \cite{Raasakka26} for more discussion on this topic.) However, the explicit analysis of these limits will have to be left for future work.

Of course, the simple model we considered above is far from realistic, and one can only hope to achieve a qualitative agreement with reality. Let us point out a couple of changes to our analysis that will follow with more realistic models:
\begin{enumerate}
    \item In 2d the gravitational constant is dimensionless. To the contrary, in 4d the volume of the faces will be proportional to $l_\Delta^2$, linking the discretization scale to the magnitude of the effective gravitational constant $G_\text{eff} \propto l_\Delta^2$. If the discretization scale is set to around Planck length, we should obtain the correct experimentally observed magnitude for $G_\text{eff}$ as in the original continuum calculation by Sakharov \cite{Sakharov68}, assuming that the proportionality constant comes out as approximately of order unity.
    \item In 2d the gravitational action (either continuum or discretized) can be reduced to a topological term, and therefore 2d gravity does not have any local physical degrees of freedom. However, the general form of the discretized gravitational (Regge) action as a sum over deficit angles around codimension-2 subsimplices remains intact in any dimension \cite{Regge61}. Likewise, in any spacetime dimension, the QFT vacuum fluctuations will give rise to particle loops around codimension-2 subsimplices, thus contributing additional Aharonov--Bohm-like phases to the quantum amplitudes in the same way as in the 2d case. Therefore, the link between the QFT vacuum fluctuations and the effective gravitational action remains intact despite the higher dimensionality of the discretized spacetime manifold.
    \item If we consider a model with several different fields, each field contributes with its own loop amplitude to the effective gravitational action.
\end{enumerate}
We leave the detailed analysis of these more realistic models also to future work.

\section{QFT vector model}
Our results in the previous section indicate that high-energy QFT vacuum fluctuations in triangulated spacetime can give rise to an effective low-energy gravitational action. If that is the case, we do not need necessarily to include gravitational dynamics explicitly into the fundamental model in order to obtain a unified theory of quantum matter and gravity. Now we proceed to apply these insights into formulating a new approach to quantum gravity.

In the previous sections we only considered situations where the triangulated background spacetime had a definite geometric structure. In order to obtain a model for quantum gravity, we need to allow for superpositions of different spacetime structures --- and eventually sum over all possible geometries, which are compatible with the given boundary data. In our framework this means that we no longer have just one tensor network for the boundary amplitude. Instead, the amplitude is obtained as a linear combination of many different tensor networks. This implies that the correlations in the boundary amplitudes no longer reflect a unique bulk spacetime structure.

The sum over tensor network amplitudes corresponding to different bulk spacetimes could be implemented as in CDT by numerically summing over the different discretizations. However, it can also be mathematically expressed using a generating function similarly to random tensor models \cite{Gurau2024} and group field theory \cite{Oriti09}. Again, for concreteness, consider the 2d scalar field model from the previous section. In the model we have two kinds of edges, spacelike and timelike, and two kinds of triangles $T$ and $T^*$ giving rise to the amplitude tensors $A^{(T)}_{{\bf n}_1 {\bf n}_2 {\bf n}_3}$ and $A^{(T^*)}_{{\bf n}_1 {\bf n}_2 {\bf n}_3}$. Denote the state space on a single spacelike edge by $\mathcal{H}_s$ and on a single timelike edge by $\mathcal{H}_t$. Now consider the action functional
\begin{align}
    S[\Psi^{(i)}_{\bf n};\lambda] &= \frac{1}{2}\sum_{i=s,t} G^{-1}_{{\bf n}_1 {\bf n}_2} \Psi^{(i)}_{{\bf n}_1} \Psi^{(i)}_{{\bf n}_2} + \frac{\lambda}{2} \sum_{j=T,T^*} A^{(j)}_{{\bf n}_1 {\bf n}_2 {\bf n}_3} \Psi^{(s)}_{{\bf n}_1} \Psi^{(t)}_{{\bf n}_2} \Psi^{(t)}_{{\bf n}_3} \,,
\end{align}
where repeated Fock space indices ${\bf n}_k$ are summed over, and $\Psi^{(i)}_{\bf n}$ ($i=s,t$) are state vectors on spacelike and timelike edges --- the dynamical variables in this action. $G^{-1}_{{\bf n}_1 {\bf n}_2}$ is the inverse gluing tensor giving rise the quadratic terms in the action. $A^{(j)}_{{\bf n}_1 {\bf n}_2 {\bf n}_3}$ ($j=T,T^*$) are the amplitude tensors corresponding to the triangles $T$ and $T^*$, which give rise to the qubic interaction terms. Finally, $\lambda$ is a coupling constant. Then, the partition function
\begin{align} \label{eq:Z}
    Z(\lambda) = \int \left[\prod_{i=s,t} \text{d}\Psi^{(i)}\right] e^{-S[\Psi^{(i)};\lambda]}
\end{align}
gives rise to a random vector model based on the previous QFT model on triangulated spacetime. This partition function defines the \emph{QFT vector model}. Similarly to random tensor models \cite{Gurau2024} and group field theory \cite{Oriti09}, the diagrams arising from the Feynman expansion of $Z(\lambda)$ in powers of $\lambda$ correspond exactly to all closed triangulated manifolds (and pseudo-manifolds), which are obtained by gluing triangles $T$ and $T^*$ together along either spacelike or timelike edges. The Feynman rules for computing the amplitudes consist of inserting one amplitude tensor for each interaction vertex (i.e., a triangle), whose indices are contracted with the gluing tensor according to the structure of the manifold. Thus, the Feynman amplitudes reproduce exactly the amplitudes for QFT in triangulated spacetimes we considered in the previous section, except weighted by $\lambda^{N_t}/\Sigma_\Delta$ where $N_t$ is the number of triangles in the triangulation and $\Sigma_\Delta$ is a symmetry factor for the Feynman diagram. Accordingly, the partition function contains the full non-perturbative model, which is obtained by summing over the amplitudes for all possible triangulated manifolds.\footnote{Although at the formal level we can define the partition function as above, its non-perturbative existence and/or the convergence of the Feynman expansion are very non-trivial properties, and will be left for future work to analyze.} Moreover, the proportionality of amplitudes to $\lambda^{N_t}$ allows to renormalize the effective cosmological constant by rescaling the QFT vector model coupling constant $\lambda$. (See Appendix \ref{app:lambda} and \cite{Raasakka26} for more discussion on the effective cosmological constant.)

Amplitudes for boundary states can be expressed as expectation values in the QFT vector model. Consider a spacetime region $R$, whose boundary $\partial R$ consists of $N$ edges. The kinematical boundary state space is then given by $\mathcal{H}_{\partial R}^\text{kin} = S(\otimes_{k=1}^N \mathcal{H}_{i_k})$, where $i_k \in \{s,t\} \forall k$ and $S$ denotes the appropriate symmetrization. Let $\chi_{{\bf n}_1 \ldots {\bf n}_N}$ be the components of a boundary state $|\chi\rangle \in \mathcal{H}_{\partial R}^\text{kin}$. Then the quantum gravitational QFT vector model amplitude for this boundary state is given by
\begin{align}
    &A(\chi_{{\bf n}_1 \ldots {\bf n}_N};\lambda) = \frac{1}{Z(\lambda)} \int \left[\prod_{i=s,t}\text{d} \Psi^{(i)}\right] \left[ \prod_{k=1}^N G_{{\bf m}_k {\bf n}_k} \right] \left[ \prod_{k=1}^N \Psi^{(i_k)}_{{\bf m}_k} \right] \chi_{{\bf n}_1 \ldots {\bf n}_N} e^{-S[\Psi^{(i)};\lambda]} \,.
\end{align}
The dynamical fields $\Psi^{(i_k)}_{{\bf m}_k}$ in the tensor contraction act as sources, coupling the boundary field $\chi_{{\bf n}_1 \ldots {\bf n}_N}$ to the bulk dynamics. The integration results in a sum over all amplitudes corresponding to bulk geometries, which are compatible with the boundary data.

It is interesting to note that the high-energy regime of the 2d scalar QFT vector model resembles random matrix models of the kind that have been previously linked to 2d quantum gravity \cite{Francesco95}. This is the regime where each boundary edge has exactly two excitations---the components $M_{kl}$ of a two-particle edge state $M_{kl}|kl\rangle \in \mathcal{H}_i$, where $|kl\rangle$ is a state with one excitation in each mode $k$ and $l$, acting as a random matrix. Intriguingly, this is also the regime where the leading order effective gravitational terms arise to the action. However, the full QFT vector model is much richer than a simple matrix model, of course, and the degrees of freedom have a clear physical interpretation unlike in many random matrix models.\footnote{In matrix models the degrees of freedom that are summed over are often considered to be just abstract matrix indices (see e.g.~\cite{Francesco95}), whereas here they are the boundary states of the matter QFT associated to the faces of the triangulation.}

QFT vector models corresponding to other QFT models and higher dimensions can also be defined in exactly the same way as in the case of 2d scalar field. In higher dimensions the UV limit would give rise to a tensor model. Controlling the sum over all triangulations is a very difficult problem. Imposing restrictions on the possible gluings (for example, as is done in colored tensor models \cite{Gurau2024}) will likely be necessary to make the perturbative expansion of the partition function well-defined. On the other hand, the QFT vector model partition function Eq.~\ref{eq:Z} may also be amenable to other forms of analysis (possibly non-perturbative in $\lambda$), e.g., via mean-field approximation, similarly to recent explorations in the context of group field theory \cite{Marchetti23}.

\section{Summary and conclusions}
To conclude, let us summarize our results. We first developed a theoretical framework, based on compositional QFT and tensor network methods, for studying QFT in triangulated spacetime. We then showed that the high-energy vacuum fluctuations in free massive scalar QFT in 2d triangulated spacetime give rise to an effective low-energy gravitational action, and discussed the possible generalization of these results to more realistic models. Finally, we introduced the QFT vector model as a unified model for quantum matter and gravity, in which the effective quantum gravitational amplitudes arise from the high-energy vacuum fluctuations of the matter fields, thus removing the need to explicitly introduce gravitational dynamics. Obviously, much work remains ahead to show that such a model can be mathematically well-defined or physically relevant.

Our work opens up several new avenues of research. Studying in more detail the emergence of low-energy effective gravitational dynamics in more realistic QFT models in 4d will be the necessary first step in bridging the gap to reality. However, the 2d QFT vector model also warrants further investigation. It would be interesting to see if the successes of 2d matrix models and CDT in reproducing reasonable macroscopic spacetime geometries can be achieved also in this model. Other interesting questions include, e.g., the fate of unitarity in non-trivial and superposed spacetime geometries. Finally, the exact formulation and study of QFT vector models for more realistic QFT models (e.g., incorporating fermions and local gauge symmetry) is essential to see if the approach can in fact successfully merge QFT and gravity.

\section*{Acknowledgments}
We acknowledge the computational resources provided by the Aalto Science-IT project.

\appendix

\section{Numerical evaluation of single-particle process matrices}\label{app:Bs}
Here we present the explicit numerical evaluation of the single-particle process matrices $B^{(ij)}$ for free massive scalar field theory in 2d Lorentzian spacetime.

We consider two oppositely oriented flat Lorentzian equilateral triangles $T$ and $T^*$, as explained in the main text. (See Fig.\ \ref{fig:lorentziantriangles}.) The boundary of $T$ consists of three boundary edges $e_i\subset \partial T$, $i=1,2,3$, so that $\partial T = \cup_i e_i$, and similarly for $T^*$. Each edge has proper length/time $l_\Delta > 0$. The intersections of the edges are the vertices of the triangle, $e_i\cap e_j = v_{k}$, $k=1,2,3$ s.t. $k\neq i,j$. The points on the edges of $T$ are given by the 2-vectors $x^{(i)} = s t^{(i)} + l_\Delta c^{(i)}$, where $s\in [-l_\Delta/2, l_\Delta/2]$ parametrizes the points, $t^{(i)}$ is the tangent vector to the edge $(i)$, and $l_\Delta c^{(i)}$ is the center of the edge $(i)$. Explicitly, we choose the triangle to have its base (edge $(1)$) on the $x$-axis and be symmetric with respect to the $t$-axis, giving
\begin{align}
    & t^{(1)} = (0, 1)\,,\qquad\qquad\quad   c^{(1)} = (0, 0)  \,,\\
    & t^{(2)} = (\sqrt{5}/2, -1/2)\,,\quad\   c^{(2)} = (\sqrt{5}/4, 1/4) \,,\\
    & t^{(3)} = (-\sqrt{5}/2, -1/2)\,,\ \  c^{(3)} = (\sqrt{5}/4, -1/4) \,.
\end{align}
The edge vectors for $T^*$ are obtained via spacetime reflection as $-x^{(i)}(s)$.

An orthogonal basis of functions on each edge is given by the functions
\begin{align*}
    \zeta_k(s) = \frac{1}{l_\Delta} e^{i 2\pi \frac{k}{l_\Delta}s} \,,\quad k\in\mathbb{Z} \,,
\end{align*}
where again $s\in [-\frac{l_\Delta}{2},\frac{l_\Delta}{2}]$ is the coordinate labeling points along the edge. (Note that $\zeta_{-k} = \overline{\zeta_k}$.) The normalization is chosen such that they satisfy the orthogonality relation
\begin{align*}
    (\zeta_k,\zeta_l) = \int_{[-\frac{l_\Delta}{2},\frac{l_\Delta}{2}]}\text{d}s\ \overline{\zeta_k(s)} \zeta_l(s) = \frac{1}{l_\Delta} \delta_{kl} \,,
\end{align*}
which approaches the Dirac delta in the continuum limit $l_\Delta\rightarrow 0$. Also, with this normalization the probability amplitudes come out dimensionless, which is important.

The causal 2-point function (i.e., the Feynman propagator) of the free self-adjoint scalar field is defined as
\begin{align}
    G_2(x,y) = -i\langle 0 | T \hat\phi(x) \hat\phi(y) |0\rangle = -i \left\{ \begin{array}{ll} \langle 0 | \hat\phi(x) \hat\phi(y) |0\rangle & \text{ if } x_0 \geq y_0, \\ \langle 0 | \hat\phi(y) \hat\phi(x) |0\rangle & \text{ if } x_0 < y_0, \end{array} \right.
\end{align}
where $T$ denotes the time-ordering. Using standard methods, this can be expressed as \cite{Huang10}
\begin{align}
    G_2(x,y) = -i \int_\mathbb{R} \frac{\text{d}p_1}{4\pi p_0} e^{-\text{sgn}(x_0-y_0) i p\cdot (x-y)} \,,
\end{align}
where $p_0 = \sqrt{p_1^2 + m^2}$, $p=(p_0,p_1)$, and `$\text{sgn}$' denotes the sign function. The elements of the single-particle process matrix $B^{(ij)}$ can then be generally expressed as
\begin{align*}
    B^{(ij)}_{kl} &= \int_{-\frac{l_\Delta}{2}}^{\frac{l_\Delta}{2}} \text{d}s \int_{-\frac{l_\Delta}{2}}^{\frac{l_\Delta}{2}} \text{d}s'\ \zeta_k(s) \zeta_l(s') G_2(x^{(i)}(s), x^{(j)}(s')) \,.
\end{align*}
Substituting the expressions for the basis functions $\zeta_k$ and the 2-point function $G_2$, the integrals over the parameters $s$ and $s'$ can be carried out analytically. We note that $B^{(ji)}_{kl} = B^{(ij)}_{lk}$. Furthermore, the neutral scalar field two-point function is symmetric under a spacetime reflection (i.e., a $PT$-transformation), $G_2(-x,-y) = G_2(x,y)$, so the amplitudes for $T$ and $T^*$ are equal.

\begin{figure}
    \centering\includegraphics[width=0.5\linewidth]{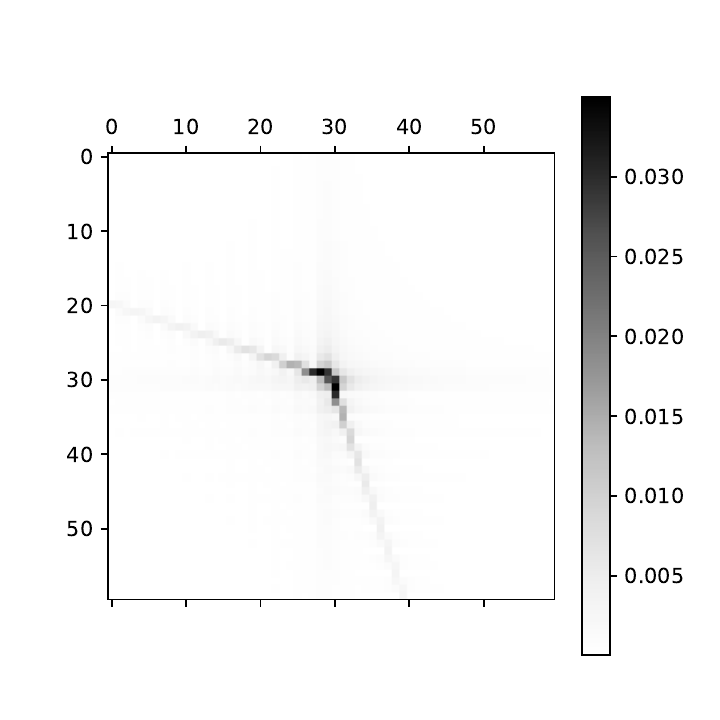}
    \caption{Numerically evaluated absolute values of the matrix elements $B^{(23)}_{kl}$ for $|k|,|l|<30$.}
    \label{fig:B23abs}
\end{figure}

The explicit expressions for the $B$ matrix elements are as follows:
\begin{align}
    B^{(1j)}_{kl} &= -i \int_\mathbb{R} \frac{\text{d}p_1}{4\pi p_0} e^{ip\cdot (c^{(1)} - c^{(j)})l_\Delta} F_k(-p\cdot t^{(1)}\frac{l_\Delta}{2}) F_l(p\cdot t^{(j)}\frac{l_\Delta}{2}) \qquad \forall j=1,2,3 \,,\\
    B^{(22)}_{kl} &= \int_\mathbb{R} \frac{\text{d}p_1}{8\pi p_0} \left[ K_{kl}^{(2)}(p) - \overline{K_{kl}^{(2)}(-p)} \right] \,,\\
    B^{(33)}_{kl} &= \int_\mathbb{R} \frac{\text{d}p_1}{8\pi p_0} \left[ K_{kl}^{(3)}(-p) - \overline{K_{kl}^{(3)}(p)} \right] \,,\\
    B^{(23)}_{kl} &= \int_\mathbb{R} \frac{\text{d}p_1}{8\pi p_0} \left[ e^{-ip\cdot (c^{(2)} - c^{(3)})l_\Delta} L_{kl}^{(23)}(p) - e^{ip\cdot (c^{(2)} - c^{(3)})l_\Delta} \overline{L_{kl}^{(23)}(-p)}  \right] \,,
\end{align}
where the various special functions are
\begin{align*}
    F_k(q) &= \frac{\sin(\pi k - q)}{\pi k - q} \,,\\
    K_{kl}^{(i)}(p) &= \frac{\delta_{k,-l} - F_k(-p\cdot t^{(i)}\frac{l_\Delta}{2}) e^{i(\pi l - p\cdot t^{(i)}\frac{l_\Delta}{2})}}{\pi l - p\cdot t^{(i)}\frac{l_\Delta}{2}} \,,\\
    L_{kl}^{(ij)}(p) &= \frac{F_{k-l}(p \cdot (t^{(i)}+t^{(j)})\frac{l_\Delta}{2}) - F_k(p\cdot t^{(i)}\frac{l_\Delta}{2}) e^{i(\pi l + p\cdot t^{(j)}\frac{l_\Delta}{2})} }{\pi l + p\cdot t^{(j)}\frac{l_\Delta}{2}} \,.
\end{align*}
At the special points $p\cdot t^{(i)}\frac{l_\Delta}{2} = \pm k, \pm l$ some of the denominators of the integrands for $B^{(ij)}_{kl}$ go to zero. Even though the integrands themselves are in fact well-behaved differentiable functions at these points, they still need to be treated separately in the numerical implementation, e.g., by setting the value of the function to the limiting value close to the special point. With that taken care of, the integrals for $B^{(ij)}_{kl}$ can be numerically evaluated using standard numerical packages. We used the `quad' integrator in the SciPy Python package \cite{scipy}. The code we used for evaluating the $B$ matrix elements can be found at \cite{git}. Evaluating one $B$ matrix up to a mode cutoff $|k|,|l|\leq 50$ on one CPU core took up to $\sim 40$ minutes, but the code could definitely be optimized further, e.g., via parallelization. See Fig.~\ref{fig:B23abs} for an example of the numerically evaluated matrix elements.

\section{Discrete gravity action}\label{app:regge}
Next we describe some details of the discretization of GR following mostly the exposition in \cite{Sorkin75}. The continuum Einstein-Hilbert action in $d$-dimensional Lorentzian spacetime with metric signature $(+,-,\ldots,-)$ reads
\begin{align*}
    S_{EH}[g] = \frac{1}{2\kappa} \int_\mathcal{M} \text{d}^dx \sqrt{|\det g|} (R - 2\Lambda) \,,
\end{align*}
where $\kappa = 8\pi G/c^4$, $G$ is the gravitational constant, $\Lambda$ the cosmological constant, $R$ is the (Ricci) scalar curvature, and $g$ denotes the spacetime metric. We integrate over the whole spacetime manifold $\mathcal{M}$. (Here we assume for simplicity that $\mathcal{M}$ does not have a boundary. In the case of a non-empty boundary the action would contain additional boundary terms.) In the $d=2$ case, we discretize spacetime by considering triangulations with flat triangles $T$ and $T^*$ as discussed in the main text. Curvature of the triangulated manifold is concentrated on codimension-2 subsimplices, which in the 2d case are the 0-dimensional vertices. A closed loop around a single vertex is timelike, and the non-trivial holonomy around the loop is a 2d Lorentz boost given by the coordinate transformation
\begin{align*}
    t' &= \cosh(\theta_v)t - \sinh(\theta_v)x \,,\\
    x' &= \cosh(\theta_v)x - \sinh(\theta_v)t \,,
\end{align*}
where $\theta_v$ is a hyperbolic deficit angle, which quantifies the curvature at the vertex $v$ \cite{Regge61,Sorkin75}. The Einstein-Hilbert action for the triangulated manifold can then be written as
\begin{align*}
    S_{EH} = \frac{1}{\kappa} \left[ \sum_v V_v \theta_v - V_\Delta \Lambda \right] \,,
\end{align*}
where $V_v$ is the volume of a single vertex, and $V_\Delta$ is the spacetime volume of the triangulation. We set $V_v\equiv 1$ following the convention in \cite{Ambjorn01}. For the total spacetime volume of the triangulation we obviously have $V_\Delta = N_t V_t$, where $N_t$ is the number of triangles in the triangulation and $V_t$ the volume of a single triangle.

In two spacetime dimensions the integral over the curvature leads in fact to a purely topological term \cite{Loll15}, which would allow us to further simplify the form of the action. However, we keep the curvature term explicit in our analysis, since this is the form that arises naturally from the virtual particle loops. Moreover, it also helps us to see how our results in 2d would generalize to higher dimensions, where GR is no longer topological. The unsimplified form of the action generalizes in a straightforward manner to higher dimensions, where the sum still runs over all the codimension-2 subsimplices with deficit angles quantifying the curvature \cite{Regge61}.

\section{Cosmological constant}\label{app:lambda}
The main challenge for the original induced gravity proposal of Sakharov \cite{Sakharov68} was to reproduce the observed value for the cosmological constant \cite{Visser02}. Indeed, as is well-known, the effective cosmological constant arising from continuum QFT vacuum fluctuations comes out as dozens of orders of magnitude too large --- the infamous `worst prediction in the history of physics'. Therefore, it is crucial to examine if the framework we introduced can help with this problem.

There are at least two ways we can get around the problem: (i) By considering microscopic QFT models for which the contributions to the effective cosmological constant from bosonic and fermionic degrees of freedom (nearly) cancel out, since these contribute with opposite signs. (ii) As noted in the main text, the effective cosmological constant can be renormalized in the QFT vector model by rescaling the QFT vector model coupling constant $\lambda$, since the QFT vector model Feynman amplitudes are proportional to $\lambda^V$, where $V$ is the total spacetime volume. The running of $\lambda$ as a function of the renormalization scale can then be studied. This cures the cosmological constant problem in the QFT vector model. However, further investigation is certainly needed in order to fill in the technical details.

With the above general remarks in mind, in the rest of this section we concentrate on the numerical evaluation of the leading order contributions to the effective cosmological constant in the 2d free scalar field model. These arise from virtual single-particle loops crossing over only one edge. The contribution to the amplitude coming from a loop crossing the spacelike edge (1) is real-valued, so it does not contribute to the effective gravitational action. Both timelike edges (2) and (3) contribute to the amplitude a factor
\begin{align}
    \text{tr}(\tilde G \tilde B^{(ii)} \tilde G \tilde B^{(ii)}) \propto e^{i\gamma} \,,
\end{align}
where $\gamma \approx 3.1$. Since each edge is shared by two triangles, and each triangle has two timelike edges, this is also the leading order contribution by one triangle to the effective cosmological constant. The cosmological constant term in the gravitational action can be written as
\begin{align}
    e^{\frac{i}{\kappa}\Lambda V_tN_t} = \prod_{t=1}^{N_t} e^{\frac{i}{\kappa}\Lambda V_t} \,,
\end{align}
where the product runs over all the triangles in the triangulation. Thus, equating the factor $e^{\frac{i}{\kappa}\Lambda V_t}$ arising from one triangle with the phase of the loop amplitude $e^{i\gamma}$, we get for the effective cosmological constant $\Lambda_\text{eff} = \frac{\kappa_\text{eff}}{V_t}\gamma$, where $\kappa_\text{eff} = 8\pi G_\text{eff}$ and $G_\text{eff} = \frac{\theta_1}{8\pi\Delta\varphi} \approx 0.015$ is the effective gravitational constant evaluated in the main text. The volume of a single triangle ($T$ or $T^*$) is given by $V_t = \frac{\sqrt{5}}{2}l_\Delta^2$. Substituting everything in, we get $\Lambda_\text{eff} = \frac{16\pi G_\text{eff}\gamma}{\sqrt{5}} l_\Delta^{-2} \approx 1.0\ l_\Delta^{-2}$. Thus, the effective cosmological constant comes out the same order of magnitude as $l_\Delta^{-2}$, and diverges in the continuum limit $l_\Delta \rightarrow 0$ --- as could have been expected from the continuum theory.


\end{document}

%% file: tensor_contract.eps_tex
\begingroup%
  \makeatletter%
  \providecommand\color[2][]{%
    \errmessage{(Inkscape) Color is used for the text in Inkscape, but the package 'color.sty' is not loaded}%
    \renewcommand\color[2][]{}%
  }%
  \providecommand\transparent[1]{%
    \errmessage{(Inkscape) Transparency is used (non-zero) for the text in Inkscape, but the package 'transparent.sty' is not loaded}%
    \renewcommand\transparent[1]{}%
  }%
  \providecommand\rotatebox[2]{#2}%
  \newcommand*\fsize{\dimexpr\f@size pt\relax}%
  \newcommand*\lineheight[1]{\fontsize{\fsize}{#1\fsize}\selectfont}%
  \ifx\svgwidth\undefined%
    \setlength{\unitlength}{544.04794287bp}%
    \ifx\svgscale\undefined%
      \relax%
    \else%
      \setlength{\unitlength}{\unitlength * \real{\svgscale}}%
    \fi%
  \else%
    \setlength{\unitlength}{\svgwidth}%
  \fi%
  \global\let\svgwidth\undefined%
  \global\let\svgscale\undefined%
  \makeatother%
  \begin{picture}(1,0.31712535)%
    \lineheight{1}%
    \setlength\tabcolsep{0pt}%
    \put(0,0){\includegraphics[width=\unitlength]{tensor_contract.eps}}%
    \put(0.02666913,0.17197941){\color[rgb]{0,0,0}\makebox(0,0)[lt]{\lineheight{1.25}\smash{\begin{tabular}[t]{l}$\mathbf{n}_1$\end{tabular}}}}%
    \put(0.17610948,0.07074326){\color[rgb]{0,0,0}\makebox(0,0)[lt]{\lineheight{1.25}\smash{\begin{tabular}[t]{l}$\mathbf{n}_2$\end{tabular}}}}%
    \put(0.1864515,0.19149484){\color[rgb]{0,0,0}\makebox(0,0)[lt]{\lineheight{1.25}\smash{\begin{tabular}[t]{l}$\mathbf{n}_3$\end{tabular}}}}%
    \put(0.10717018,0.17441887){\color[rgb]{0,0,0}\makebox(0,0)[lt]{\lineheight{1.25}\smash{\begin{tabular}[t]{l}$A$\end{tabular}}}}%
    \put(0.66797149,0.12928958){\color[rgb]{0,0,0}\makebox(0,0)[lt]{\lineheight{1.25}\smash{\begin{tabular}[t]{l}$A$\end{tabular}}}}%
    \put(0.7911623,0.08813701){\color[rgb]{0,0,0}\makebox(0,0)[lt]{\lineheight{1.25}\smash{\begin{tabular}[t]{l}$A$\end{tabular}}}}%
    \put(0.90337585,0.22564688){\color[rgb]{0,0,0}\makebox(0,0)[lt]{\lineheight{1.25}\smash{\begin{tabular}[t]{l}$A$\end{tabular}}}}%
    \put(0.75335123,0.16710072){\color[rgb]{0,0,0}\makebox(0,0)[lt]{\lineheight{1.25}\smash{\begin{tabular}[t]{l}$G$\end{tabular}}}}%
    \put(0.86890628,0.1341685){\color[rgb]{0,0,0}\makebox(0,0)[lt]{\lineheight{1.25}\smash{\begin{tabular}[t]{l}$G$\end{tabular}}}}%
    \put(0.43654357,0.18539643){\color[rgb]{0,0,0}\makebox(0,0)[lt]{\lineheight{1.25}\smash{\begin{tabular}[t]{l}$G$\end{tabular}}}}%
    \put(0.38929259,0.15856254){\color[rgb]{0,0,0}\makebox(0,0)[lt]{\lineheight{1.25}\smash{\begin{tabular}[t]{l}$\mathbf{n}_1$\end{tabular}}}}%
    \put(0.4813554,0.15278166){\color[rgb]{0,0,0}\makebox(0,0)[lt]{\lineheight{1.25}\smash{\begin{tabular}[t]{l}$\mathbf{n}_2$\end{tabular}}}}%
  \end{picture}%
\endgroup%

%% file: lorentzian_triangles.eps_tex
\begingroup%
  \makeatletter%
  \providecommand\color[2][]{%
    \errmessage{(Inkscape) Color is used for the text in Inkscape, but the package 'color.sty' is not loaded}%
    \renewcommand\color[2][]{}%
  }%
  \providecommand\transparent[1]{%
    \errmessage{(Inkscape) Transparency is used (non-zero) for the text in Inkscape, but the package 'transparent.sty' is not loaded}%
    \renewcommand\transparent[1]{}%
  }%
  \providecommand\rotatebox[2]{#2}%
  \newcommand*\fsize{\dimexpr\f@size pt\relax}%
  \newcommand*\lineheight[1]{\fontsize{\fsize}{#1\fsize}\selectfont}%
  \ifx\svgwidth\undefined%
    \setlength{\unitlength}{630.26535875bp}%
    \ifx\svgscale\undefined%
      \relax%
    \else%
      \setlength{\unitlength}{\unitlength * \real{\svgscale}}%
    \fi%
  \else%
    \setlength{\unitlength}{\svgwidth}%
  \fi%
  \global\let\svgwidth\undefined%
  \global\let\svgscale\undefined%
  \makeatother%
  \begin{picture}(1,0.35782205)%
    \lineheight{1}%
    \setlength\tabcolsep{0pt}%
    \put(0,0){\includegraphics[width=\unitlength]{lorentzian_triangles.eps}}%
    \put(0.09848175,0.13987925){\color[rgb]{0,0,0}\makebox(0,0)[lt]{\lineheight{1.25}\smash{\begin{tabular}[t]{l}$x$\end{tabular}}}}%
    \put(0.0348814,0.21168291){\color[rgb]{0,0,0}\makebox(0,0)[lt]{\lineheight{1.25}\smash{\begin{tabular}[t]{l}$t$\end{tabular}}}}%
    \put(0.39874653,0.22365969){\color[rgb]{0,0,0}\makebox(0,0)[lt]{\lineheight{1.25}\smash{\begin{tabular}[t]{l}$l_\Delta$\end{tabular}}}}%
    \put(0.24588352,0.02993956){\color[rgb]{0,0,0}\makebox(0,0)[lt]{\lineheight{1.25}\smash{\begin{tabular}[t]{l}$l_\Delta$\end{tabular}}}}%
    \put(0.83158229,0.12885247){\color[rgb]{0,0,0}\makebox(0,0)[lt]{\lineheight{1.25}\smash{\begin{tabular}[t]{l}$l_\Delta$\end{tabular}}}}%
    \put(0.67275254,0.31984674){\color[rgb]{0,0,0}\makebox(0,0)[lt]{\lineheight{1.25}\smash{\begin{tabular}[t]{l}$l_\Delta$\end{tabular}}}}%
    \put(0.1693314,0.31427602){\color[rgb]{0,0,0}\makebox(0,0)[lt]{\lineheight{1.25}\smash{\begin{tabular}[t]{l}$T$\end{tabular}}}}%
    \put(0.56630888,0.32411847){\color[rgb]{0,0,0}\makebox(0,0)[lt]{\lineheight{1.25}\smash{\begin{tabular}[t]{l}$T^*$\end{tabular}}}}%
    \put(0.21037289,0.09375394){\color[rgb]{0,0,0}\makebox(0,0)[lt]{\lineheight{1.25}\smash{\begin{tabular}[t]{l}$(1)$\end{tabular}}}}%
    \put(0.6931666,0.25303432){\color[rgb]{0,0,0}\makebox(0,0)[lt]{\lineheight{1.25}\smash{\begin{tabular}[t]{l}$(1)$\end{tabular}}}}%
    \put(0.31928072,0.1766706){\color[rgb]{0,0,0}\makebox(0,0)[lt]{\lineheight{1.25}\smash{\begin{tabular}[t]{l}$(2)$\end{tabular}}}}%
    \put(0.5818159,0.16645135){\color[rgb]{0,0,0}\makebox(0,0)[lt]{\lineheight{1.25}\smash{\begin{tabular}[t]{l}$(2)$\end{tabular}}}}%
    \put(0.74423944,0.16593159){\color[rgb]{0,0,0}\makebox(0,0)[lt]{\lineheight{1.25}\smash{\begin{tabular}[t]{l}$(3)$\end{tabular}}}}%
    \put(0.15319127,0.17628939){\color[rgb]{0,0,0}\makebox(0,0)[lt]{\lineheight{1.25}\smash{\begin{tabular}[t]{l}$(3)$\end{tabular}}}}%
    \put(0.24315257,0.21401023){\color[rgb]{0,0,0}\makebox(0,0)[lt]{\lineheight{1.25}\smash{\begin{tabular}[t]{l}$\theta_1$\end{tabular}}}}%
    \put(0.67251614,0.12677371){\color[rgb]{0,0,0}\makebox(0,0)[lt]{\lineheight{1.25}\smash{\begin{tabular}[t]{l}$\theta_1$\end{tabular}}}}%
  \end{picture}%
\endgroup%

%% file: decomp.eps_tex
\begingroup%
  \makeatletter%
  \providecommand\color[2][]{%
    \errmessage{(Inkscape) Color is used for the text in Inkscape, but the package 'color.sty' is not loaded}%
    \renewcommand\color[2][]{}%
  }%
  \providecommand\transparent[1]{%
    \errmessage{(Inkscape) Transparency is used (non-zero) for the text in Inkscape, but the package 'transparent.sty' is not loaded}%
    \renewcommand\transparent[1]{}%
  }%
  \providecommand\rotatebox[2]{#2}%
  \newcommand*\fsize{\dimexpr\f@size pt\relax}%
  \newcommand*\lineheight[1]{\fontsize{\fsize}{#1\fsize}\selectfont}%
  \ifx\svgwidth\undefined%
    \setlength{\unitlength}{492.71774172bp}%
    \ifx\svgscale\undefined%
      \relax%
    \else%
      \setlength{\unitlength}{\unitlength * \real{\svgscale}}%
    \fi%
  \else%
    \setlength{\unitlength}{\svgwidth}%
  \fi%
  \global\let\svgwidth\undefined%
  \global\let\svgscale\undefined%
  \makeatother%
  \begin{picture}(1,0.18422976)%
    \lineheight{1}%
    \setlength\tabcolsep{0pt}%
    \put(0,0){\includegraphics[width=\unitlength]{decomp.eps}}%
    \put(0.21690391,0.07039798){\color[rgb]{0,0,0}\makebox(0,0)[lt]{\lineheight{1.25}\smash{\begin{tabular}[t]{l}$=$\end{tabular}}}}%
    \put(0.47883695,0.07344228){\color[rgb]{0,0,0}\makebox(0,0)[lt]{\lineheight{1.25}\smash{\begin{tabular}[t]{l}$\times$\end{tabular}}}}%
    \put(0.70896024,0.07344228){\color[rgb]{0,0,0}\makebox(0,0)[lt]{\lineheight{1.25}\smash{\begin{tabular}[t]{l}$\times$\end{tabular}}}}%
  \end{picture}%
\endgroup%

%% file: renorm.eps_tex
\begingroup%
  \makeatletter%
  \providecommand\color[2][]{%
    \errmessage{(Inkscape) Color is used for the text in Inkscape, but the package 'color.sty' is not loaded}%
    \renewcommand\color[2][]{}%
  }%
  \providecommand\transparent[1]{%
    \errmessage{(Inkscape) Transparency is used (non-zero) for the text in Inkscape, but the package 'transparent.sty' is not loaded}%
    \renewcommand\transparent[1]{}%
  }%
  \providecommand\rotatebox[2]{#2}%
  \newcommand*\fsize{\dimexpr\f@size pt\relax}%
  \newcommand*\lineheight[1]{\fontsize{\fsize}{#1\fsize}\selectfont}%
  \ifx\svgwidth\undefined%
    \setlength{\unitlength}{420.8541798bp}%
    \ifx\svgscale\undefined%
      \relax%
    \else%
      \setlength{\unitlength}{\unitlength * \real{\svgscale}}%
    \fi%
  \else%
    \setlength{\unitlength}{\svgwidth}%
  \fi%
  \global\let\svgwidth\undefined%
  \global\let\svgscale\undefined%
  \makeatother%
  \begin{picture}(1,0.35487527)%
    \lineheight{1}%
    \setlength\tabcolsep{0pt}%
    \put(0,0){\includegraphics[width=\unitlength]{renorm.eps}}%
    \put(0.66893422,0.04081629){\color[rgb]{0,0,0}\makebox(0,0)[lt]{\lineheight{1.25}\smash{\begin{tabular}[t]{l}$t-\Delta t$\end{tabular}}}}%
    \put(0.70861675,0.16893419){\color[rgb]{0,0,0}\makebox(0,0)[lt]{\lineheight{1.25}\smash{\begin{tabular}[t]{l}$t$\end{tabular}}}}%
    \put(0.65759634,0.29478456){\color[rgb]{0,0,0}\makebox(0,0)[lt]{\lineheight{1.25}\smash{\begin{tabular}[t]{l}$t+\Delta t$\end{tabular}}}}%
  \end{picture}%
\endgroup%

%% file: oneloop.eps_tex
\begingroup%
  \makeatletter%
  \providecommand\color[2][]{%
    \errmessage{(Inkscape) Color is used for the text in Inkscape, but the package 'color.sty' is not loaded}%
    \renewcommand\color[2][]{}%
  }%
  \providecommand\transparent[1]{%
    \errmessage{(Inkscape) Transparency is used (non-zero) for the text in Inkscape, but the package 'transparent.sty' is not loaded}%
    \renewcommand\transparent[1]{}%
  }%
  \providecommand\rotatebox[2]{#2}%
  \newcommand*\fsize{\dimexpr\f@size pt\relax}%
  \newcommand*\lineheight[1]{\fontsize{\fsize}{#1\fsize}\selectfont}%
  \ifx\svgwidth\undefined%
    \setlength{\unitlength}{460.70760412bp}%
    \ifx\svgscale\undefined%
      \relax%
    \else%
      \setlength{\unitlength}{\unitlength * \real{\svgscale}}%
    \fi%
  \else%
    \setlength{\unitlength}{\svgwidth}%
  \fi%
  \global\let\svgwidth\undefined%
  \global\let\svgscale\undefined%
  \makeatother%
  \begin{picture}(1,0.7977083)%
    \lineheight{1}%
    \setlength\tabcolsep{0pt}%
    \put(0,0){\includegraphics[width=\unitlength]{oneloop.eps}}%
    \put(0.11677368,0.56538218){\color[rgb]{0,0,0}\makebox(0,0)[lt]{\lineheight{1.25}\smash{\begin{tabular}[t]{l}$x$\end{tabular}}}}%
    \put(0.02976601,0.66361229){\color[rgb]{0,0,0}\makebox(0,0)[lt]{\lineheight{1.25}\smash{\begin{tabular}[t]{l}$t$\end{tabular}}}}%
    \put(0.4187162,0.03019779){\color[rgb]{0,0,0}\makebox(0,0)[lt]{\lineheight{1.25}\smash{\begin{tabular}[t]{l}$n_v$\end{tabular}}}}%
    \put(0.41871619,0.73784759){\color[rgb]{0,0,0}\makebox(0,0)[lt]{\lineheight{1.25}\smash{\begin{tabular}[t]{l}$n_v^*$\end{tabular}}}}%
    \put(0.40225919,0.14865239){\color[rgb]{0,0,0}\makebox(0,0)[lt]{\lineheight{1.25}\smash{\begin{tabular}[t]{l}$\tilde B^{(23)}$\end{tabular}}}}%
    \put(0.234961,0.27335523){\color[rgb]{0,0,0}\makebox(0,0)[lt]{\lineheight{1.25}\smash{\begin{tabular}[t]{l}$\tilde B^{(31)}$\end{tabular}}}}%
    \put(0.56630159,0.27309158){\color[rgb]{0,0,0}\makebox(0,0)[lt]{\lineheight{1.25}\smash{\begin{tabular}[t]{l}$\tilde B^{(12)}$\end{tabular}}}}%
    \put(0.23821688,0.49768247){\color[rgb]{0,0,0}\makebox(0,0)[lt]{\lineheight{1.25}\smash{\begin{tabular}[t]{l}$\tilde B^{(12)}$\end{tabular}}}}%
    \put(0.55732511,0.50093832){\color[rgb]{0,0,0}\makebox(0,0)[lt]{\lineheight{1.25}\smash{\begin{tabular}[t]{l}$\tilde B^{(31)}$\end{tabular}}}}%
    \put(0.40674748,0.62238522){\color[rgb]{0,0,0}\makebox(0,0)[lt]{\lineheight{1.25}\smash{\begin{tabular}[t]{l}$\tilde B^{(23)}$\end{tabular}}}}%
    \put(0.46958321,0.37676299){\color[rgb]{0,0,0}\makebox(0,0)[lt]{\lineheight{1.25}\smash{\begin{tabular}[t]{l}$v$\end{tabular}}}}%
  \end{picture}%
\endgroup%

%% file: ampang.eps_tex
\begingroup%
  \makeatletter%
  \providecommand\color[2][]{%
    \errmessage{(Inkscape) Color is used for the text in Inkscape, but the package 'color.sty' is not loaded}%
    \renewcommand\color[2][]{}%
  }%
  \providecommand\transparent[1]{%
    \errmessage{(Inkscape) Transparency is used (non-zero) for the text in Inkscape, but the package 'transparent.sty' is not loaded}%
    \renewcommand\transparent[1]{}%
  }%
  \providecommand\rotatebox[2]{#2}%
  \newcommand*\fsize{\dimexpr\f@size pt\relax}%
  \newcommand*\lineheight[1]{\fontsize{\fsize}{#1\fsize}\selectfont}%
  \ifx\svgwidth\undefined%
    \setlength{\unitlength}{413.64916386bp}%
    \ifx\svgscale\undefined%
      \relax%
    \else%
      \setlength{\unitlength}{\unitlength * \real{\svgscale}}%
    \fi%
  \else%
    \setlength{\unitlength}{\svgwidth}%
  \fi%
  \global\let\svgwidth\undefined%
  \global\let\svgscale\undefined%
  \makeatother%
  \begin{picture}(1,0.75407098)%
    \lineheight{1}%
    \setlength\tabcolsep{0pt}%
    \put(0,0){\includegraphics[width=\unitlength]{ampang.eps}}%
    \put(0.14950989,0.0529972){\color[rgb]{0,0,0}\makebox(0,0)[lt]{\lineheight{1.25}\smash{\begin{tabular}[t]{l}1\end{tabular}}}}%
    \put(0.24761675,0.0529972){\color[rgb]{0,0,0}\makebox(0,0)[lt]{\lineheight{1.25}\smash{\begin{tabular}[t]{l}2\end{tabular}}}}%
    \put(0.34572361,0.0529972){\color[rgb]{0,0,0}\makebox(0,0)[lt]{\lineheight{1.25}\smash{\begin{tabular}[t]{l}3\end{tabular}}}}%
    \put(0.44383044,0.0529972){\color[rgb]{0,0,0}\makebox(0,0)[lt]{\lineheight{1.25}\smash{\begin{tabular}[t]{l}4\end{tabular}}}}%
    \put(0.54193729,0.0529972){\color[rgb]{0,0,0}\makebox(0,0)[lt]{\lineheight{1.25}\smash{\begin{tabular}[t]{l}5\end{tabular}}}}%
    \put(0.64004415,0.0529972){\color[rgb]{0,0,0}\makebox(0,0)[lt]{\lineheight{1.25}\smash{\begin{tabular}[t]{l}6\end{tabular}}}}%
    \put(0.73815101,0.0529972){\color[rgb]{0,0,0}\makebox(0,0)[lt]{\lineheight{1.25}\smash{\begin{tabular}[t]{l}7\end{tabular}}}}%
    \put(0.83625786,0.0529972){\color[rgb]{0,0,0}\makebox(0,0)[lt]{\lineheight{1.25}\smash{\begin{tabular}[t]{l}8\end{tabular}}}}%
    \put(0.93436472,0.0529972){\color[rgb]{0,0,0}\makebox(0,0)[lt]{\lineheight{1.25}\smash{\begin{tabular}[t]{l}9\end{tabular}}}}%
    \put(0.5419562,0.01994534){\color[rgb]{0,0,0}\makebox(0,0)[lt]{\lineheight{1.25}\smash{\begin{tabular}[t]{l}$n$\end{tabular}}}}%
    \put(0.05532552,0.1466363){\color[rgb]{0,0,0}\makebox(0,0)[lt]{\lineheight{1.25}\smash{\begin{tabular}[t]{l}2.6\end{tabular}}}}%
    \put(0.05532552,0.24943369){\color[rgb]{0,0,0}\makebox(0,0)[lt]{\lineheight{1.25}\smash{\begin{tabular}[t]{l}2.7\end{tabular}}}}%
    \put(0.05532552,0.35223109){\color[rgb]{0,0,0}\makebox(0,0)[lt]{\lineheight{1.25}\smash{\begin{tabular}[t]{l}2.8\end{tabular}}}}%
    \put(0.05532552,0.45502849){\color[rgb]{0,0,0}\makebox(0,0)[lt]{\lineheight{1.25}\smash{\begin{tabular}[t]{l}2.9\end{tabular}}}}%
    \put(0.05532552,0.55782587){\color[rgb]{0,0,0}\makebox(0,0)[lt]{\lineheight{1.25}\smash{\begin{tabular}[t]{l}3.0\end{tabular}}}}%
    \put(0.05532552,0.66062327){\color[rgb]{0,0,0}\makebox(0,0)[lt]{\lineheight{1.25}\smash{\begin{tabular}[t]{l}3.1\end{tabular}}}}%
    \put(0.03700534,0.38555876){\color[rgb]{0,0,0}\rotatebox{90}{\makebox(0,0)[lt]{\lineheight{1.25}\smash{\begin{tabular}[t]{l}$\Delta\varphi$\end{tabular}}}}}%
  \end{picture}%
\endgroup%